\documentclass{aa}  

\usepackage[utf8]{inputenc}
\usepackage{pdfpages}
\usepackage[colorlinks=True,allcolors=blue,final]{hyperref}
\usepackage{ragged2e,array,url,graphicx,csquotes,adjustbox,longtable}
\usepackage[english]{babel}
\usepackage{caption}
\usepackage{xcolor}
\usepackage{amsmath}
\usepackage{subcaption}
\usepackage{natbib}
\usepackage{aas_macros}
\usepackage{parskip}
\usepackage{float}
\usepackage{indentfirst}
\usepackage{enumitem}
\usepackage[title,titletoc]{appendix}
\usepackage{colortbl}
\usepackage{threeparttable}
\usepackage{tocloft} 
\usepackage{booktabs}
\usepackage{bm}
\usepackage{comment}
\usepackage[normalem]{ulem}
\usepackage{makecell} 

\begin{document}

   \title{Towards a definition of a meteor cluster}

   \subtitle{Detection of meteor clusters from meteor orbit databases}

   \author{A. Ashimbekova
          \inst{1}
          \and
          J. Vaubaillon\inst{1}
          \and
          P. Koten\inst{2}
          }

   \institute{LTE, Observatoire de Paris, Université PSL, Sorbonne Université, Université de Lille, LNE, CNRS, 61 Avenue de l’Observatoire, 75014 Paris, France\\
              \email{aishabibi.ashimbekova@obspm.fr}
            \and
            Astronomical Institute of the Czech Academy of Sciences, Fricova 298, 25165, Ond\v{r}ejov, Czech Republic
             }

   \date{}

 
  \abstract
   {As of today, there is no official definition of a meteor cluster. It is usually identified as a large number of meteors sharing a similar radiant and velocity, all occurring within a few seconds. Only eight clusters {have been} reported so far, from single-camera or camera network observations. However, a cluster may be observed by several distant networks and remain unnoticed simply because each network is recording a small portion of the cluster.} 
   {We aim to provide an overview of meteor clusters to help define what constitutes a cluster by potentially adding more to the already identified ones and determining their common parameters.}
   {A search for new clusters is performed in publicly available International Astronomical Union (IAU) meteor databases with the DBSCAN algorithm. Then, a statistical significance method is applied to derive the most promising cluster candidates. However, the method still lacks a way to debias the atmospheric area surveyed by the cameras due to a lack of publicly available data.}
   {A set of 16 statistically significant potential clusters is identified, involving 4 to 7 fragments. The 90th percentile includes a duration of 8 seconds, a velocity difference of 2.2 km/s, and a radiant spread of nearly 4 degrees. The velocity difference may arise from the method used for orbit computation.}
   {Meteor clusters might be more frequent than currently reported. However, we recommend that future meteor orbit databases also include a way to estimate the surveyed area by the cameras involved in the detection. This would strengthen the veracity of the 16 identified cluster candidates and ultimately allow scientists to fully debias the number of clusters, and hence derive the physical lifetime expectancy of meteoroids, which is often overlooked due to the focus on collisional lifetime estimates only. {We also recommend that any future cluster observation report include the expected number of random occurrences and consider the event to be real if this value is below 0.1.}}


   \maketitle
%

\section{Introduction}\label{sec:intro}

Meteor showers are characterized by a relatively high number of meteors sharing a similar radiant and occurring over a time span ranging from a few hours to a few weeks \citep{Koten2019,Williams2019}.
During such showers, meteor clusters are rarely seen. 
A meteor cluster is characterized by a relatively high number of meteors sharing similar radiant and velocity, occurring within a time interval of a few seconds.
Until today, only 8 clusters were reported \citep{Hapgood1981_c1,Piers1993_c2,Kinoshita1999_c3,Watanabe2002_c4,Watanabe2003_c5,Koten2017_c6,Vaubaillon2023_c7,Koten2024_c8}. 
However, the recognition of clusters is impaired by the lack of an official definition, caused by the scarce frequency of the phenomenon.
Another cause of the absence of further detection of meteor clusters is the meteor detection software, not explicitly designed for this purpose \citep{Bednar2023}.

Yet, the existence of meteor clusters is an indicator of meteoroid fragmentation in interplanetary space.
The most likely physical origin is the thermal stress endured by the meteoroid \citep{Capek2022_physclust}.
Determining the frequency of such self-fragmentation directly impacts estimates of meteoroid physical lifetimes in interplanetary space.
In particular, if the frequency is found to be higher than already suspected, this could explain, e.g., the absence of the 2006 Leonids outburst \citep{Jenniskens2008}.
This physical lifetime expectancy is often overlooked, in favor of collisional lifetime expectancy \citep{Koschny2017,Soja2019}.
Thus, determining the frequency of meteoroid self-fragmentation by the detection of meteor clusters has implications for the interpretation of meteor shower observations in general.
The scientific question this paper seeks to address is whether there are more meteor clusters hidden in the meteor orbit databases than are currently known.

Recent search for additional meteor pairs among the Geminids recorded by the Ond\v{r}ejov observatory video meteor database \citep{Koten2021} was performed by \cite{Durisova2023}. 
The results show that no cluster was detected, given the data and the statistical significance of the pair candidates.
This statistical analysis is crucial, since, as pointed out by the authors, a simple Poisson distribution underestimates the expected number of clusters in a given database of meteor orbits.

However, all currently known clusters and search for clusters focus on detection with a limited number of cameras (often restricted to just one).
The interpretation \citep{Capek2022_physclust} is that a meteoroid self-fragments in space, and the subsequent non-gravitational forces (highly dependent on the size) cause a slow increase of the physical distance between the different fragments.
The measured maximal physical distance between the fragments constrain the maximum time duration between the self-fragmentation of the observation in the Earth's atmosphere.
It is found that, for the analyzed clusters, the maximal physical distance was estimated at a few hundred km, and the corresponding duration since self-fragmentation does not exceed a few days at most \citep{Koten2017_c6,Koten2024_c8}.

However, the studied clusters were detected either by a single camera or by a network of cameras, surveying a rather small portion of the atmosphere.
Now let us suppose a simple scenario of a meteoroid self-fragmentation into two pieces (i.e., the most basic cluster possible).
Suppose the self-fragmentation happens a few months before the fragments enter the Earth's atmosphere and the physical distance is 3000~km at the time of the observation.
Such a distance would make it impossible for this cluster to be detected by a two-camera-network (since cameras of a given network are usually separated by $\sim 100 \; km$).
Even a network of 100 cameras covering $10^6  \ km^2$ cannot detect such a cluster.
However, the two meteors can be detected by two meteor observation networks a few thousand km apart.
In this case, the two meteors associated with the atmospheric entry of the two fragments will be detected, but the link between these two meteors cannot be performed, unless the data are available.

The goal of this paper is to dig into the current publicly available databases, find potential meteor clusters, and perform a statistical analysis to evaluate their respective relevance.
Since there is no current definition of a meteor cluster, it may be tempting to define constraints to help identify future clusters.
However, such a procedure might unintentionally set too strong limits, preventing future works to provide scientifically sound results.
Instead, we aim to provide preliminary statistics of the cluster candidates.
It is worth mentioning that the consequences regarding the physical lifetime expectancy are out of the scope of this preliminary study, since this is the topic of thorough research.

The paper first describes the method in section \ref{sec:method} (including the description of the databases and the algorithm).
Then, the cluster candidates are characterized in section \ref{sec:res} and a general discussion on future considerations regarding the definition of a meteor cluster is included in section \ref{sec:discuss}.

\section{Method}\label{sec:method}

The overall method described below is voluntarily conservative in order to only detect the most likely clusters.
In particular, we were reluctant to define any strict criteria out of concern for introducing bias in the analysis.

\subsection{Meteor orbit databases}
Most of the teams that have detected meteor clusters have utilized their local datasets. On the contrary, our objective for this project was to combine multiple large databases over a larger time interval. 
To begin with, we chose to work with the International Astronomical Union Meteor Data Center (IAU MDC) video database\footnote{https://iaumeteordatacenter.org, accessed in Jan. 2024} as it is thoroughly documented. 
IAU MDC is composed of two meteor orbit databases, SonotaCo and CAMS.

\subsubsection{SonotaCo}
SonotaCo is a database of several hundred thousand meteor orbits from automated multi-station video observations over Japan \citep{SonotaCo2009, SonotaCo2021}. The network has began operating on January 1, 2007 and has since observed the night sky above Japan without interruptions. There have been 20 to 30 stations across Japan during this period with about 429 registered cameras, a quarter of which have observed continuously. Typical cameras are hi-sensitivity monochrome CCD video cameras with a field of view of 30-90\textdegree. SonotaCo utilizes a self-developed motion detection software called UFOCapture that enables video recording starting a few seconds before the trigger \citep{SonotaCo2009}. 

\subsubsection{CAMS}
CAMS stands for Cameras for Allsky Meteor Surveillance and is an automated video surveillance operating in California, USA \citep{CAMS2011}. The network for which data are available in the IAU database surveys the night sky from three stations in northern California, located at Fremont Peak Observatory, Lick Observatory, and at a site in Mountain View. There are sixty video cameras in total, all above 31\textdegree \ elevation. All cameras are identical and have a narrow-angle field-of-view of 30\textdegree. CAMS utilizes detection algorithms and modules from the MeteorScan software package for automatic processing of meteor data. Nowadays{,} CAMS includes {many} more stations and cameras spread out {across} the world.

The total temporal extent of the IAU MDC database is from January 1, 2007 to December 31, 2022 (16 years), with the SonotaCo dataset covering the entire time span, while the CAMS dataset covering only 7 years (from January 1, 2010 to December 31, 2016). It should be noted that the already observed clusters are not included in the SonotaCo and CAMS databases.

\subsection{Clustering algorithm (DBSCAN)}
Given the vastness of our dataset, it was important to develop an efficient way to detect potential meteor clusters. The clustering algorithm called DBSCAN, developed by \cite{Ester1996_DBSCAN}, was selected as the most suitable approach.
DBSCAN stands for Density-based Spatial Clustering of Applications with Noise\footnote{\hyperlink{https://scikit-learn.org/stable/modules/generated/sklearn.cluster.DBSCAN.html}{Documentation on sklearn.cluster.DBSCAN (accessed in February 2024)}} and it is an algorithm within the machine learning library of Python, scikit-learn. This algorithm was chosen for fulfilling requirements that are crucial when working with large spatial databases: 1) minimal requirements of domain knowledge to determine the input parameters, as suitable values are often not known in advance when dealing with large databases; 2) ability to detect clusters of arbitrary shapes and varying densities; 3) good efficiency on databases of significantly more than just a few thousand objects \citep{Ester1996_DBSCAN}. DBSCAN was previously tested and utilized in the context of meteors by \cite{Sugar2017}, but for the detection of meteor showers. According to their findings, the method proved to be highly efficient.

DBSCAN takes in two parameters as arguments: epsilon ($\epsilon$), which defines the radius within which to search for neighboring data points; and minPoints, the minimum number of points to form a cluster. The algorithm works by assigning all data points into three categories: core points, border points, and noise points (outliers). Core points are those that have at least minPoints within their epsilon neighborhood, border points have fewer than minPoints within their epsilon neighborhood but are in the neighborhood of a core point, and noise points have neither enough points within their epsilon neighborhood nor are they in the epsilon neighborhood of a core point. This process ensures that all points are examined and clusters are expanded from any core point found during iterations \citep{Ester1996_DBSCAN}. 

\subsubsection{Parameter space}\label{sec:param}
We apply the clustering algorithm to a 4-dimensional parameter space composed of the observable geocentric quantities of the meteoroids:
\begin{equation}
\label{eq:4dvector}
\vec{v} = \begin{bmatrix}
t \\
v_g \\
\alpha_g \\
\delta_g
\end{bmatrix}
\end{equation}
where $t$ is the observation time of the meteor, $v_g$ is the geocentric velocity, and $\alpha_g$ and $\delta_g$ are the equatorial coordinates of the geocentric radiant given by the right ascension (RA) and declination (DEC), respectively. We normalize all 4 parameters so that the influence of each variable is comparable when determining clusters. It is important to note that meteor clusters occur within a narrow time frame, often within a few seconds. Consequently, small differences in time are crucial for accurately identifying and grouping cluster members. To address this sensitivity, we normalize the time parameter to a range of [$-10^5$,$10^5$] instead of the [$-$1,1] range used for the other parameters. Empirical testing showed that normalizing all parameters to the same range ([$-$1,1]) indeed resulted in poor clustering performance. By expanding the normalization range of the time parameter, we effectively increase its weight, thus achieving more meaningful clustering results across all parameters. 

\subsubsection{Distance criterion}
One of the two arguments that DBSCAN requires the user to define is the distance criterion, $\epsilon$, as previously mentioned. It represents the maximum distance within which data points can be considered neighbors. For each core point, the algorithm includes all points in the $\epsilon$-neighborhood in the cluster (i.e., points whose distance from the core point is less than or equal to epsilon). 

The distances are computed as the Euclidean norm of the difference between two meteor vectors and is conceptually similar to the D-parameter defined by \cite{Valsecchi1999}. However, the norm offers the advantage of a faster computation due to the built-in functionality provided by the scikit-learn package \citep{Sugar2017}. 

\begin{figure}[h]
    \centering
    \includegraphics[width=0.40\textwidth]{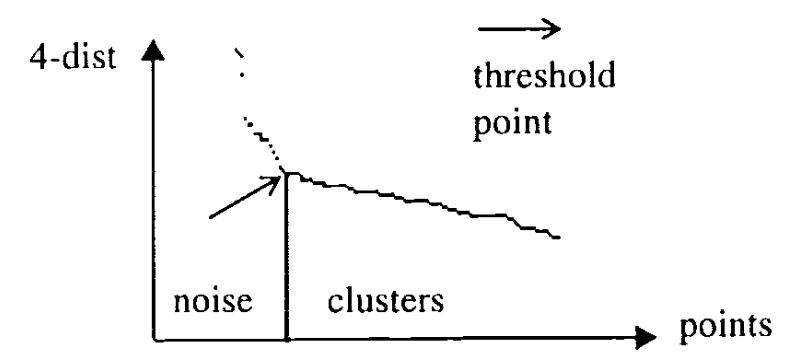}
    \caption{Graph of distance of a point to its $4^{th}$ nearest neighbor (sorted in descending order) for a sample database in \cite{Ester1996_DBSCAN}.}
    \label{fig:bend_dbscan}
\end{figure}

In order for the algorithm to perform efficiently, it is important to choose an optimal value for epsilon. The developers of DBSCAN suggest to use the following method to determine this parameter: let $d$ be a distance of a point $p$ to its $k^{th}$ nearest neighbor. For a given $k$ define a function $k-$dist, mapping each point to the distance from its $k^{th}$ nearest neighbor and sorting them in descending order. The graph of this function is believed to provide some clues on the density distribution of the data points. The threshold point (i.e., the optimal epsilon value) is then the first point of the first ``valley'' of the sorted $k-$dist graph as shown in Figure \ref{fig:bend_dbscan}. Points left of the threshold would be considered to be noise and the ones right of the threshold would be assigned to some cluster \citep{Ester1996_DBSCAN}. 

\cite{Sugar2017} also employs a similar approach by constructing a sorted nearest-neighbor distance plot. They propose two options for determining epsilon: {either by identifying} the point of maximum curvature from the plot or selecting a value corresponding to the expected percentage of clustered meteors. They chose the latter approach, determining through experimentation that 23\% of the observed meteors in their dataset are affiliated with meteor showers. We opted to experiment with the distance to nearest neighbor method, but considering the vastness of our dataset and the relative rarity and lower density of meteor clusters compared to meteor showers, the ratio of core points to noise points would make graphical detection impractical. Consequently, we do not consider this method robust enough or scientifically meaningful for our study. Additionally, we were hesitant to employ the expected percentage of clustered meteors approach as it involves making assumptions. While it may be suitable for showers, given their well-documented nature and ample amount of data, clusters lack sufficient data for us to make any assumptions confidently. Hence, having not discovered a significant approach for selecting the epsilon value, we decided to experiment with different values (``growing epsilon method''), to then prioritize the most promising clusters based on the characteristics of the output parameters (further detailed in Section \ref{sec:characteristics}).

\begin{figure}[h]
    \centering
    \includegraphics[width=0.45\textwidth]{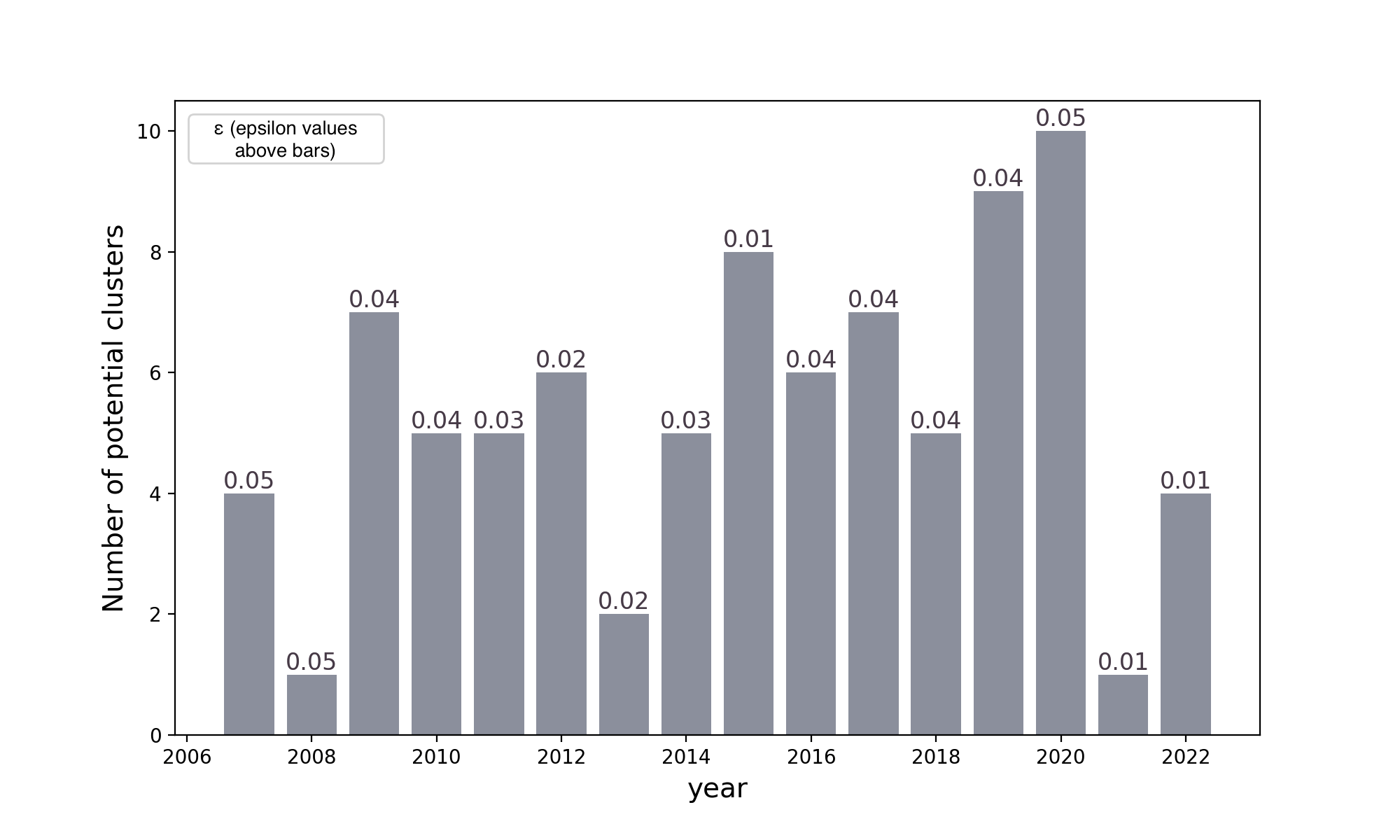}
    \caption{Distribution of clusters and corresponding $\epsilon$ values {(shown above each bar)} per year.}
    \label{fig:e_distr}
\end{figure}

For each year within the dataset, the algorithm was applied using a range of epsilon values from 0.01 to 0.05, with increments of 0.01. In prioritizing the identification of most favorable clusters, a criterion was set to select an epsilon value within the range that yielded at most 10 clusters. This approach ensures that the resulting clusters are composed of the ``closest'' possible neighboring meteors. Even a slightly larger $\epsilon$ value produces significantly more clusters with greater differences in parameters. This is not desirable for our purposes since we know that cluster meteors typically appear within seconds of each other with very similar speeds and directions. Consequently, the number of potential clusters analyzed varies per year but is always 10 or fewer. The chosen $\epsilon$ values as well as the resulting number of clusters per year are summarized in Figure \ref{fig:e_distr}. 

\section{Results}\label{sec:res}

\subsection{First set of potential clusters}\label{sec:characteristics}
Applying the clustering algorithm to the entire dataset, we identified {an initial set} of 85 potential meteor clusters over a span of 16 years (2007-2022).
It is worth mentioning at this point that this first set does not take into account the statistical significance tests, described in sec. \ref{sec:statanal}.

\begin{figure}[h]
    \centering
    \includegraphics[width=0.49\textwidth]{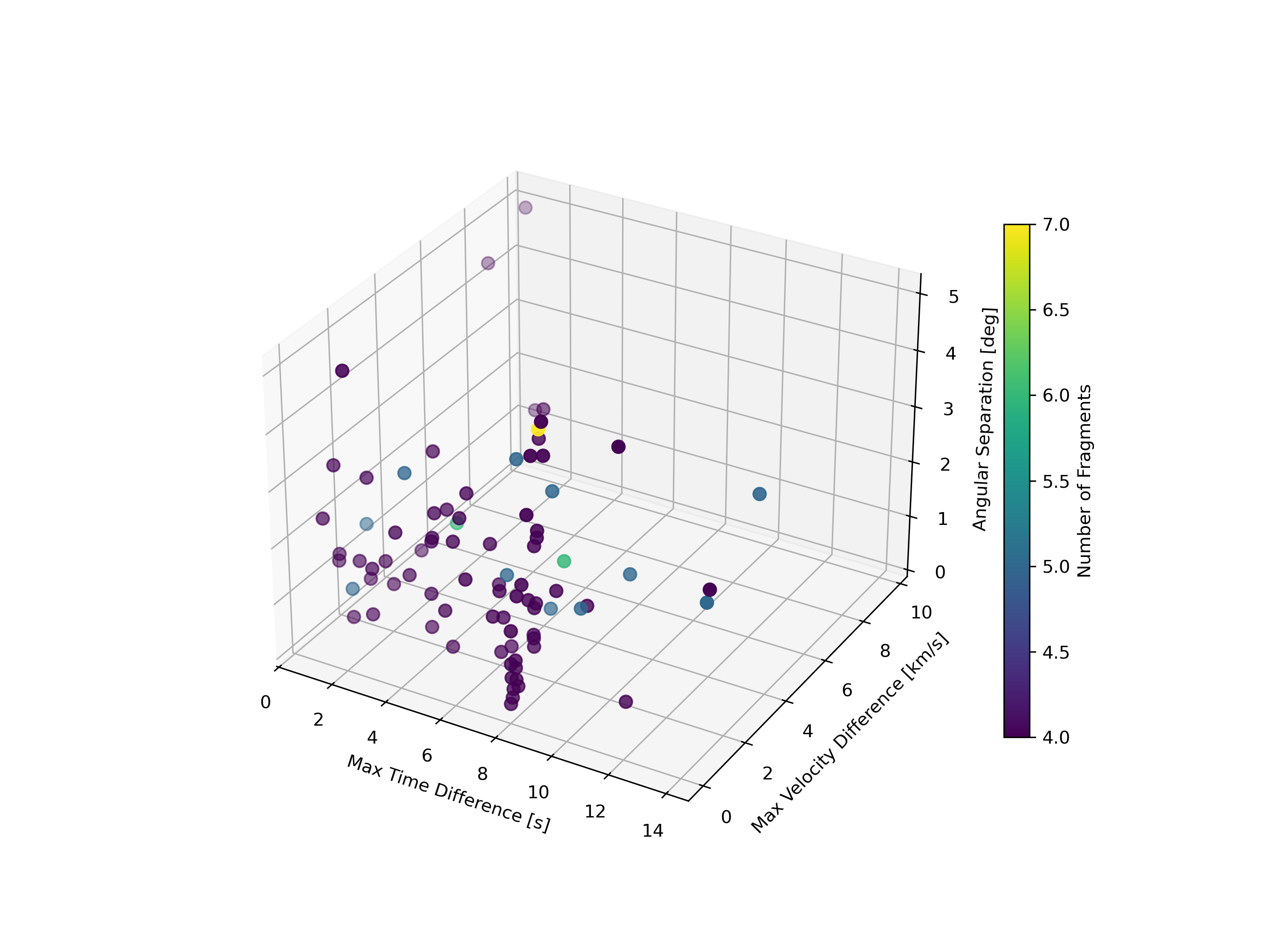}
    \caption{Parameter characteristics of the first set of cluster candidates (i.e., before the study on the statistical significance).}
    \label{fig:3dplot}
\end{figure}

We examined the parameters of the identified clusters such as the number of fragments within a cluster, the maximum time separation between member fragments, their maximum velocity separation, and maximum angular separation between their radiants. These characteristics are summarized in Figure \ref{fig:3dplot}. With the limited data available on confirmed clusters, we do not yet have well-defined upper limits for these quantities. Consequently, it is challenging to determine precise thresholds for the maximum permissible time between meteors, or the velocity and angular separation required to be considered a cluster. 

\subsection{Shower association}\label{sec:shower association}
Having identified the potential candidates for clusters, the next question to answer is - are the clusters real or did they occur by chance? As a first step, we carried out a shower association test to determine which clusters are part of established meteor showers. In order to do so, we have utilized the distance function $D_N$ defined by \cite{Valsecchi1999} to test the orbital similarity of meteoroid orbits with the shower orbits, based on their observed geocentric quantities. 
The computation was performed for every combination of meteors from our clusters and the known meteor showers (in the IAU Meteor Shower Databse\footnote{\hyperlink{https://www.ta3.sk/IAUC22DB/MDC2022/Roje/roje_lista.php?corobic_roje=1&sort_roje=0}{IAU MDC established meteor showers list (accessed in April 2024).}}). We identified the minimum $D_N$ value for each meteor, which indicates the closest matching shower orbit. Thus, each meteor is associated with one shower based on this minimum \(D_N\) value. 

However, this association alone does not confirm that a meteor belongs to the identified shower. To determine if a given cluster is actually part of a shower, further analysis is required. Specifically, we take the median of the $D_N$ values for meteors within each cluster. According to \cite{Shober2024}, a $D_N$ criterion value of $\sim$~0.15 or less achieves a false-positive shower identification rate of less than 5\%. Therefore, if the median $D_N$ value for a cluster is less than 0.15, we can confidently state that the cluster is part of the corresponding shower. Conversely, if the median \(D_N\) exceeds 0.15, the cluster is likely not associated with the identified shower orbit and is therefore considered part of the sporadic meteors. 

\begin{figure}[ht]
    \centering
    \includegraphics[width=0.49\textwidth]{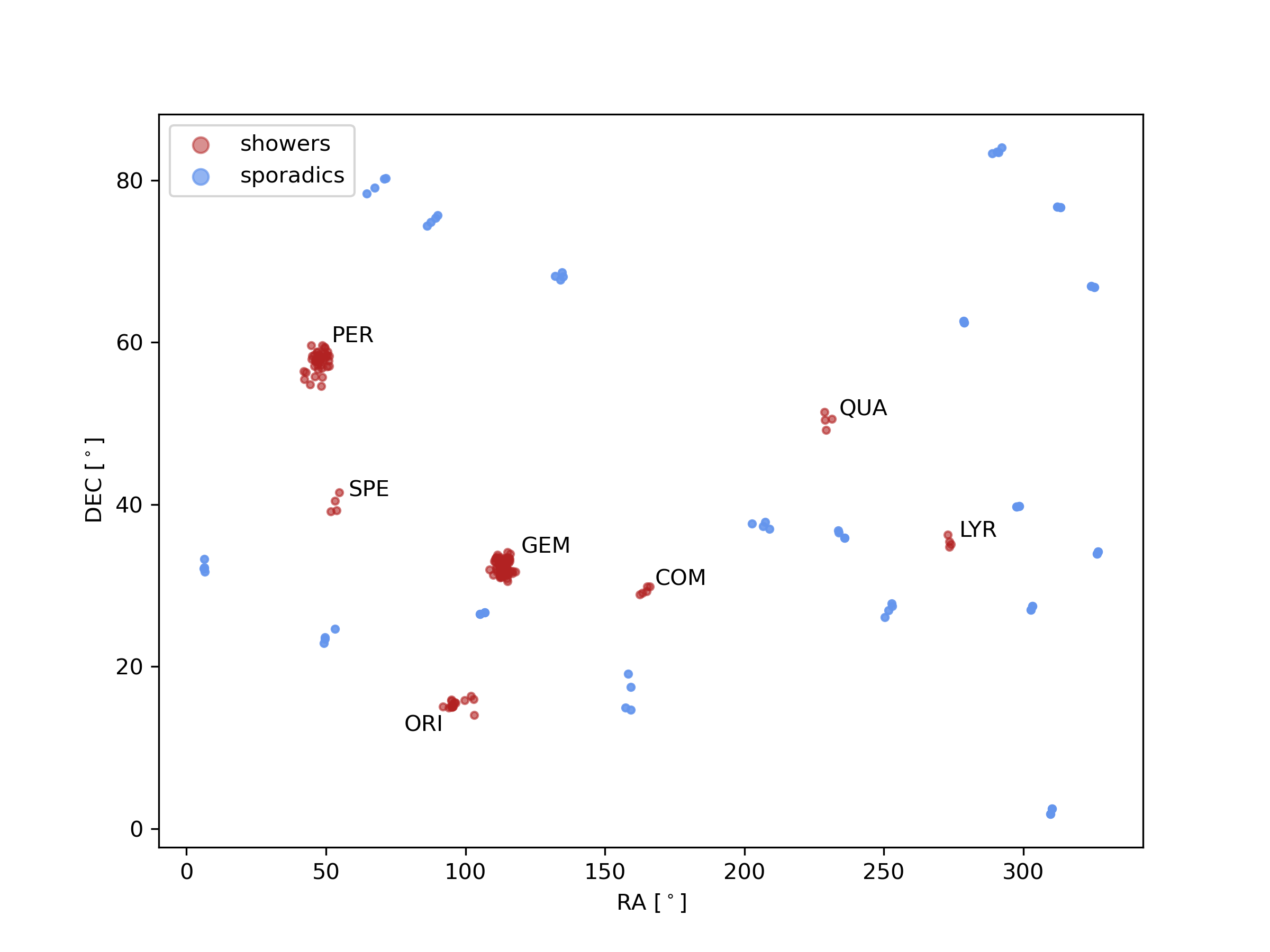}
    \caption{Association of cluster candidates with known showers and the sporadic background.}
    \label{fig:showerassoc}
\end{figure}

We found that 67 out of the 85 identified potential clusters are associated with some established showers. Figure \ref{fig:showerassoc} shows all cluster candidates indicating their shower association. In particular, 67 cluster candidates were associated with seven established showers with the majority (49 clusters) being in Geminids (GEM), 10 in Perseids (PER), 4 in Orionids (ORI), one in Quadrantids (QUA), one in Comae Berenicids (COM), one in April Lyrids (LYR), and one in September $\epsilon$-Perseids (SPE). Perseids, Geminids, and Quadrantids are among the most abundant known showers.

\subsection{Statistical significance analysis}\label{sec:statanal}
Knowing the shower associations of our identified cluster candidates, we proceeded to distinguish real clusters from chance occurrences. To do so, we compared the number of observed meteors in a cluster with a given time separation with the expected number of random appearances, $N$, that is given by the Poisson distribution \citep{Porubcan2002}:
\begin{equation}
    N = n \frac{\mu^x}{x!} e^{-\mu},
\end{equation}
where $n$ is the number of intervals (per hour), $\mu$ is the mean rate of meteors per interval, and $x$ is the number of meteors in the cluster. More specifically, if all meteors in a cluster appeared within 2~s, then $n = 1800$. Consequently, $\mu$ is calculated as follows:
\begin{equation}
    \mu = \frac{\text{ZHR}}{n},
\end{equation}
where ZHR is the zenithal hourly rate.

For each shower, we took the maximum ZHR value which corresponds to the peak activity period of a meteor shower in order to obtain the upper limit of $N$. 
Similarly, for cluster candidates not associated with showers, we took the highest observed hourly rate of the sporadic meteors, HR$_{spo} = 11.6$ \citep{Dubietis2010}. The ZHR values used in the computation of $N$s are given in Table \ref{table: ZHR} and were taken from the International Meteor Organization\footnote{\hyperlink{https://www.imo.net/files/meteor-shower/cal2024.pdf}{IMO 2024 Meteor Shower Calendar (accessed in May 2024).}} (IMO) database.

It is important to highlight that the Poisson distribution was found to significantly underestimate the number of expected clusters \citep{Koten2021}. More precisely, \cite{Sampson2007} determined from numerical analysis that the number of clusters appears to be three times more than what the Poisson model estimates. Therefore, a factor of three was added in our computations of $N$ to account for this underestimation. In fact, conversely, our computations slightly overestimate the number of expected clusters since we are using the maximum ZHR values (and not the ZHR at the time of the cluster), and the identified clusters were not necessarily observed during the peak activity of the showers or the sporadic background.
These considerations make the method rather conservative.

The ZHR is defined for a field of view of a human naked eye, which, in turn, can be converted into a surveyed atmospheric area.
Obviously, the number of observed meteors increases if this area increases.
Stricto sensu, the analysis should take this effect into account by comparing the total area surveyed by the cameras involved in detecting a potential cluster to the area visible to the naked eye.
Unfortunately, and as discussed in further detail in sec. \ref{sec:choices}, the  database does not include any information regarding the camera parameters and pointing direction to estimate this effect.
As a consequence, any cluster passing the statistical significance test described here must still be considered as a potential cluster.

\begin{table*}[t]
\centering
\caption{Maximum ZHR value and total duration (in hours) of meteor showers used in the computation of the expected number of clusters.}
\label{table: ZHR}
\begin{tabular}{l|c|c|c|c|c|c|c}
\textbf{Meteor shower} & GEM & PER & ORI & QUA & COM & LYR & SPE \\
\hline
\textbf{ZHR$_{max}$ $[h^{-1}]$} & 150 & 100 & 20 & 80 & 3 & 18 & 8 \\
\hline
\textbf{Duration $[h]$} & 96 & 168 & 168 & 24 & 24 & 24 &24 \\
\end{tabular}
\end{table*}

\begin{figure}[!htbp]
    \centering
    \includegraphics[width=0.499\textwidth]{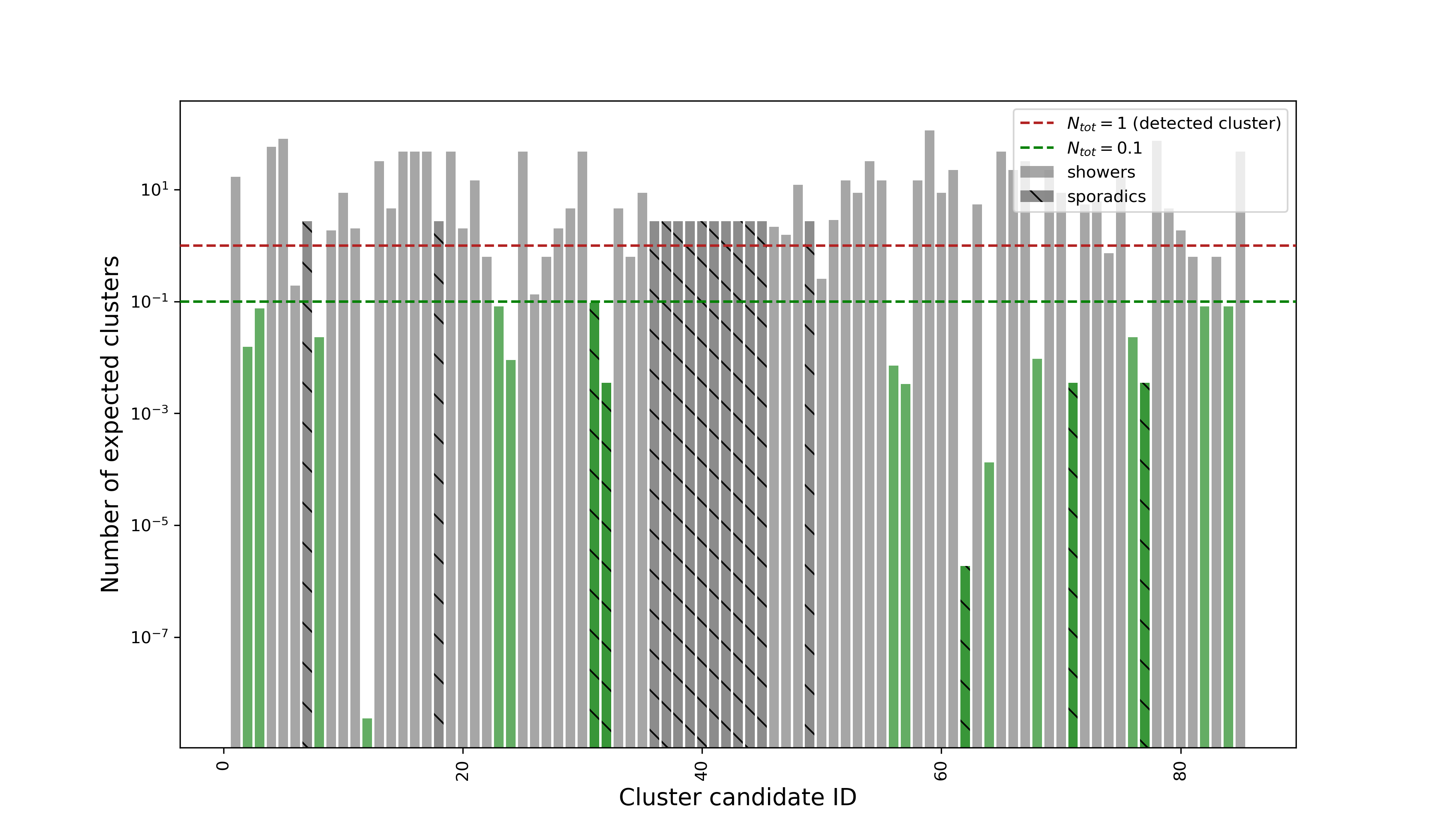}
    \caption{{For each detected cluster, this plot presents the expected number of random occurrences of such a cluster over the entire observation period. The clusters marked in green are considered potential candidates, while those in gray are dismissed as random occurrences.}}
    \label{fig:probabilities}
\end{figure}

The computed $N$ values represent the number of expected clusters for one hour of observation by naked eye. Given that the observations were continuous, we needed to account for the total duration of the meteor showers for the cluster candidates that are associated with showers. Therefore, the $N$ values were multiplied by the duration of the corresponding showers, yielding $N_{tot}$ values. For each meteor shower, we took the ``extreme'' window of time to ensure that $N$ was maintained at the upper limit. The values used are given in Table~\ref{table: ZHR}. For the cluster candidates coming from the sporadic background, we used the temporal extent of the entire dataset ($\approx 10^5$ hours) for the duration. For each cluster we have identified, the expected number of such clusters (with the same number of meteors within the same period of time) $N_{tot}$ is shown in Figure~\ref{fig:probabilities}.  We observe that, for the majority of clusters, the number of expected clusters {is significantly high, $N_{tot}>1$}, indicating that these clusters could easily occur by chance. This validates the method of tuning epsilon to identify only 10 most promising clusters per year and not more because even among these, the majority are already being dismissed as random occurrences. Therefore, our focus shifts to clusters where the number of expected clusters is significantly below one. These clusters represent instances where, given the specific number of fragments and time separation, we expect practically zero randomly occurring clusters. The fact that we detect a cluster in such a case implies that it is highly likely to be a real meteor cluster. 

To restate our approach, we restricted our focus to 10 or fewer clusters annually, utilized the maximum ZHR values to compute the expected cluster count, and set an extended observation period for the showers — all aimed at ensuring a ``conservative'' method. By adopting these measures, which involve somewhat arbitrary choices, we are deliberately going for a cautious approach. This means we are potentially overestimating the likelihood that identified clusters may not be real. This cautious methodology allows for margin of error in our selected parameters, thereby facilitating the identification of the most probable meteor clusters within the database.

Out of the initial 85 identified candidates, our statistical analysis resulted in 18 clusters with $N_{tot}$ values sufficiently low to indicate that they are highly unlikely to have occurred by chance, given the method applied here.

\subsubsection{Monte Carlo simulation}
To confirm the results of the adjusted Poisson distribution test, a Monte Carlo test for random occurrence of the meteor groups was prepared. For each cluster candidate, ZHR number of random times of meteor occurrences within one hour were generated. It was then checked whether $N$ meteors (number of meteors in the cluster) occurred within $\Delta t$ seconds (duration of the cluster). Each step was repeated 100,000 times and the sum of positive occurrences was counted. All the accepted cluster candidates passed this test at the 3-sigma confidence level. For example, cluster candidate ID 9 (see Table \ref{table:realclusters}), consisting of 7 meteors within 7.78~s, belonging to the Geminid meteor shower, produces a random appearance in only 0.018\% of all runs. The result for the ORI cluster candidate (ID 1) of four meteors observed in 6.048~s was 0.014\%. On the other hand, a candidate consisting of 4 GEM meteors observed in 6.9~s was rejected as it randomly appears in 34.5\% of all runs. Another GEM cluster of 5 meteors within 6.9~s occurred randomly in 3\% of all the runs. The latter would be acceptable at the 2-sigma level but not at the 3-sigma level. Both of these clusters were equivalently dismissed as random occurrences based on the Poisson test. Some of the results are presented in Table \ref{tab:montecarlo}. It can be concluded that the Monte Carlo test confirms the results of the adjusted Poisson test.

\begin{table*}[h]
    \centering
    \caption{Monte Carlo test results for some cluster candidates.}
    \begin{tabular}{ccccccccc}
        \toprule
        \makecell{Cluster \\ ID} & \makecell{\# of \\ meteors} & Shower & $\Delta t$ (s) & ZHR & \makecell{\# of \\ simulations} & \makecell{\# of \\ appearances} & \makecell{Probability \\ of random \\ appearance (\%)} & \makecell{3-sigma \\ confidence \\ level} \\
        \midrule
        1  & 4 & ORI & 6.048 & 20  & 100,000 & 14   & 0.014  & Accepted \\
        5  & 4 & GEM & 0.864 & 150 & 100,000 & 97   & 0.097  & Accepted \\
        6  & 5 & GEM & 1.728 & 150 & 100,000 & 11   & 0.011  & Accepted \\
        7  & 4 & spo & 2.592 & 12  & 100,000 & 0    & 0      & Accepted \\
        9  & 7 & GEM & 7.78  & 150 & 100,000 & 18   & 0.018  & Accepted \\
        13 & 6 & GEM & 4.32  & 150 & 100,000 & 21   & 0.021  & Accepted \\
        -  & 4 & GEM & 6.9   & 120 & 10,000  & 3453 & 34.53  & Rejected \\
        -  & 5 & GEM & 6.9   & 120 & 10,000  & 301  & 3.01   & Rejected \\
        -  & 4 & PER & 7.8   & 100 & 10,000  & 1177 & 11.77  & Rejected \\
        \bottomrule
    \end{tabular}
    \label{tab:montecarlo}
\end{table*}

\subsection{Potential high-confidence clusters}

The characteristics of the potential high-confidence clusters are given in Table \ref{table:realclusters}. Firstly, we rule out clusters number 14 and 16 as real meteor clusters due to the significant velocity separation $\Delta v_g$ between their member fragments, which indicates that it is implausible for these meteors to be physically related. This is because even a difference of a few $km/s$ can drastically change the orbital elements of the body, especially the semi-major axis. Checking the semi-major axes of the member meteors, we indeed find that they have a large range: 4--90~AU for cluster~14 and -2--17~AU for cluster~16 (where a negative semi-major axis represents a hyperbolic orbit with an eccentricity greater than one). However, it should be noted that SonotaCo's reduction method and computation of orbital elements may cause a high uncertainty on the semi-major axis affecting the accuracy of these values. Nevertheless, the significant variation implies that the constituent meteors of these cluster candidates most likely do not follow the same orbit. Consequently, we are left with 16 candidates that can be considered high-confidence meteor clusters.

It is also important to note that after propagating the errors, we observe that some cluster candidates exhibit exceptionally high uncertainties. For instance, the angular separations for clusters ID 4, 7, 8, and 9 are particularly large, which affects our confidence in the reliability of these clusters. While we maximize the use of the available data, this highlights the limitations of exploiting databases with high uncertainties and potential for improvement in future analysis with higher-precision data.

\begin{table*}[h]
\centering
\caption{Characteristics of the high-confidence meteor cluster candidates.} 
\label{table:realclusters}
\begin{tabular}{ccrcrlcr}
\toprule
\textbf{ID} & \textbf{\# frag.} & $\mathbf{\Delta t}$ [s] & $\mathbf{\Delta v_g}$ [km/s] & $\mathbf{\theta}$ [\textdegree] & \textbf{Shower} & \textbf{Orbit} & $\mathbf{N_{tot}}$ \\
\midrule
1 & 4 & 6.048 & 1.926$\pm$1.641 & 1.378$\pm$1.549 & ORI & HTC & 1.54e-02 \\
2 & 4 & 10.368 & 1.867$\pm$2.227 & 4.427$\pm$0.792 & ORI & HTC & 7.58e-02 \\
3 & 4 & 6.912 & 0.594$\pm$0.194 & 1.406$\pm$0.828 & ORI & HTC & 2.29e-02 \\
4 & 5 & 8.000 & 1.932$\pm$2.850 & 3.308$\pm$14.297 & COM & Unknown & 3.53e-09 \\
5 & 4 & 0.864 & 0.960$\pm$1.438 & 3.068$\pm$1.090 & GEM & Asteroid & 8.10e-02 \\
6 & 5 & 1.728 & 0.622$\pm$1.044 & 1.104$\pm$0.632 & GEM & Asteroid & 9.00e-03 \\
7 & 4 & 2.592 & 0.510$\pm$0.949 & 1.633$\pm$10.264 & Sporadic & -- & 9.50e-02 \\
8 & 4 & 0.864 & 0.440$\pm$0.481 & 2.299$\pm$20.745 & Sporadic & -- & 3.54e-03 \\
9 & 7 & 7.776 & 1.612$\pm$3.658 & 4.446$\pm$5.648 & GEM & Asteroid & 7.17e-03 \\
10 & 4 & 8.000 & 1.091$\pm$3.196 & 1.548$\pm$0.957 & LYR & LPC & 3.32e-03 \\
11 & 5 & 0.864 & 2.253$\pm$2.351 & 1.587$\pm$0.978 & Sporadic & -- & 1.87e-06 \\
12 & 4 & 8.000 & 0.787$\pm$0.068 & 3.282$\pm$2.537 & SPE & LPC/HTC? & 1.32e-04 \\
13 & 6 & 4.320 & 2.190$\pm$3.236 & 2.129$\pm$2.400 & GEM & Asteroid & 9.47e-03 \\
\sout{14} & \sout{4} & \sout{0.864} & \sout{\textcolor{red}{9.678$\pm$3.241}} & \sout{4.994$\pm$11.438} & Sporadic & -- & 3.54e-03 \\
15 & 4 & 6.912 & 2.706$\pm$1.197 & 3.812$\pm$1.997 & ORI & HTC & 2.29e-02 \\
\sout{16} & \sout{4} & \sout{0.864} & \sout{\textcolor{red}{7.902$\pm$0.885}} & \sout{4.507$\pm$1.473} & Sporadic & -- & 3.54e-03 \\
17 & 4 & 0.864 & 1.075$\pm$1.569 & 1.448$\pm$3.017 & GEM & Asteroid & 8.10e-02 \\
18 & 4 & 0.864 & 1.051$\pm$2.241 & 1.334$\pm$0.460 & GEM & Asteroid & 8.10e-02 \\
\bottomrule
\end{tabular}
\begin{tablenotes}
\item Note:  ID: cluster identification number, \# frag.: number of fragments, $\Delta t$: duration, $\Delta v_g$: maximum difference in geocentric velocity, $\theta$: maximum difference in radiant location, Shower: meteor shower linked to the cluster, Orbit: orbit type, $N_{tot}$: total number of expected clusters at the time of the detection. Strike through clusters were ruled out as high-confidence cluster candidates as $\Delta v_g$ is significantly high.
\end{tablenotes}
\end{table*}

\begin{figure}[h]
    \centering
    \includegraphics[width=0.499\textwidth]{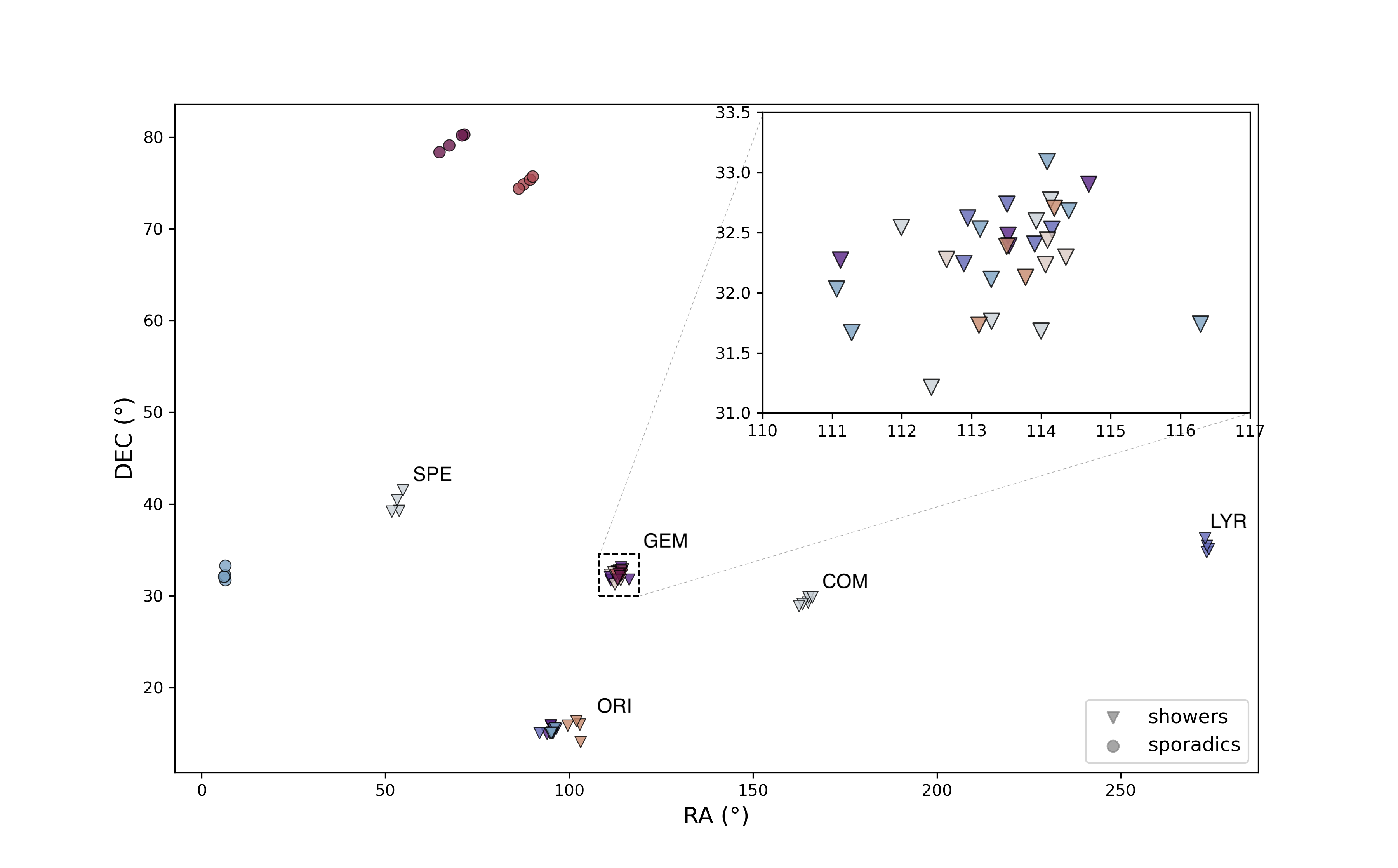}
    \caption{High-confidence meteor cluster candidates.}
    \label{fig:realclusters}
\end{figure}

Overall, the analysis reveals a blend of various meteor showers alongside sporadic occurrences for the association of the high-confidence clusters. Specifically, six clusters belong to the Geminids, four to the Orionids, one to the Comae Berenicids, one to the April Lyrids, one to the September-epsilon Perseids, and three to the sporadic background (see Figure \ref{fig:realclusters}). A single cluster was previously identified during the September-epsilon Perseids by \cite{Koten2017_c6}, with no detection recorded for the other mentioned showers \citep{Koten2021} or the sporadic background. Moreover, most of the previously detected meteor clusters have a cometary origin, whereas we observe that over one-third of our detected clusters are in Geminids whose parent body is an active asteroid, 3200 Phaethon. This is interesting because while active asteroids also contain ice, they tend to have a higher proportion of rocky material compared to cometary meteoroids. Therefore, they tend to be less porous and more cohesive. Based on the detected clusters, initially thought to originate solely from comets and thus considered more prone to self-fragmentation, it now appears that meteoroids originating from asteroids may also exhibit this behavior. 

\section{Discussion}\label{sec:discuss}

\subsection{Limitations of the method and choices}\label{sec:choices}

The main flaw in the described method is the absence of an estimate of the surveyed atmospheric area.
The computation of such area was last described in \cite{Vaubaillon2021}, but dates back to the 1990s and was improved several times \citep{KoschackRendtel90b,KoschnyZender00,GuralJenniskens00,Ocana.et.al2019}.
It involves the knowledge of the camera's field of view, but also its pointing direction above the local horizon.
The equivalent surveyed area takes into account the magnitude loss as a function of air mass, physical distance and meteor angular velocity.
The larger the surveyed area by a network (or networks), the more likely it is to detect meteors per unit of time.
This effect increases the number of expected clusters and could shorten the list of 16 identified statistically significant clusters.
This is the reason why we call them ``potential statistically significant clusters''.
To account for this effect, the effective surveyed atmosphere area should be computed over time for each camera used requiring data on field of view, pointing direction and star limiting magnitude.
Such information is routinely computed for astro-photometry and should be easily shareable by camera operators.
Ideally, any cloud cover influencing the limiting magnitude and operational interruptions should also be considered to assess performance variations.

Currently, networks estimating meteoroid mass flux compute these parameters, but they may not be available for all networks. Unless an alternative method is developed to debias the cluster count, implementing this correction remains challenging. 
However, the IAU may require the addition of the camera field of view, pointing direction and limiting magnitude (under clear sky conditions) to help in this matter.
Such parameters are nearly constant over time and would allow us to put a lower limit on the real number of clusters, and hence, help debias the observations.

\subsection{Physical origin of cluster candidates}

The physical origin of meteor clusters was thoroughly explored by \cite{Capek2022_physclust}.
They considered fast rotation, collision and thermal stress.
The latter is considered as the most plausible, although an impact scenario cannot be totally ruled out.
Considering a zero ejection velocity, dynamical analysis based on gravitational and non-gravitational forces acting on meteoroids have shown that the age of observed clusters range from a few hours to a few days at most \citep[see Fig 4 in][]{Koten2017_c6}.
The main reason is the differential non-gravitational forces acting on different sizes causing the fragment to disperse along the main orbit.
However, the physical distance between the fragments is in the order of a few hundred km at most.
Here, the fragments of the potential statistically significant clusters detected both by CAMS and SonotaCo were potentially separated by a few thousand km.
A simple extrapolation from \cite{Koten2017_c6} would conclude that their ages range from a few days to a few weeks.
However, such considerations do not take into account any possible ejection velocity caused by thermal stress or collision.
In the latter case, the physical survival of the fragments would be highly challenged, whereas for thermal stress their survival is evident.
In order to produce a physical distance between the fragments of a few thousand km, as would be the case if a cluster was detected e.g. from both Japan (SonotaCo) and California (CAMS), the ejection velocity would be evaluated from the total duration of the cluster.
Considering a physical distance of $10^5$ km and an age of 2 weeks, the estimated ejection velocity would be $\sim 8 \; m.s^{-1}$.
But such a computation ignores the role of non-gravitational forces.
Assuming that they account for only half of the physical distance of the fragments, the ejection velocity would be of a few $m.s^{-1}$ only, which is compatible with thermal stresses.

It is worth pointing out that a significant fraction of cluster candidates originate from an asteroid (3200 Phaethon).
The Geminids meteor shower is nearly as active as the Perseids, for which we find five fewer cluster candidates.
As a first approximation, we expected fragile cometary meteoroids to self-fragment more often than less fragile asteroid dust, as Geminids are known for \citep{Henych2024}.
However, our results show that clusters may occur regardless of the physical origin of the meteoroid.
The low perihelion distance of the Geminids plays a crucial role in changing the meteoroid physical structure because of the high temperature difference they experience during their orbital revolution.
For all these reasons, thermal stress appears to be the most likely physical process for the formation of meteor clusters \citep{Capek2022_physclust}.

\subsection{Towards a definition of a meteor cluster}
As previously mentioned, an official definition of a meteor cluster has not yet been established, with only eight reported clusters known to date. This study takes a step towards formulating a working definition by identifying twice as many potential cluster candidates as the currently confirmed ones. In this section, we discuss the features of these meteor clusters to help quantitatively characterize the phenomenon. However, our intention is not to set strict limits on any parameters, in order not to introduce any bias in future studies. Similarly, any strong limits, such as on the number of fragments may prove inadequate for future observations, potentially leading to false claims of meteor cluster detections when they are merely coincidental observations. Instead, we present some statistics of our set of 16 potential high-confidence meteor clusters and compare with the confirmed clusters where applicable.

\begin{figure}[h]
    \centering
    \includegraphics[width=0.45\textwidth]{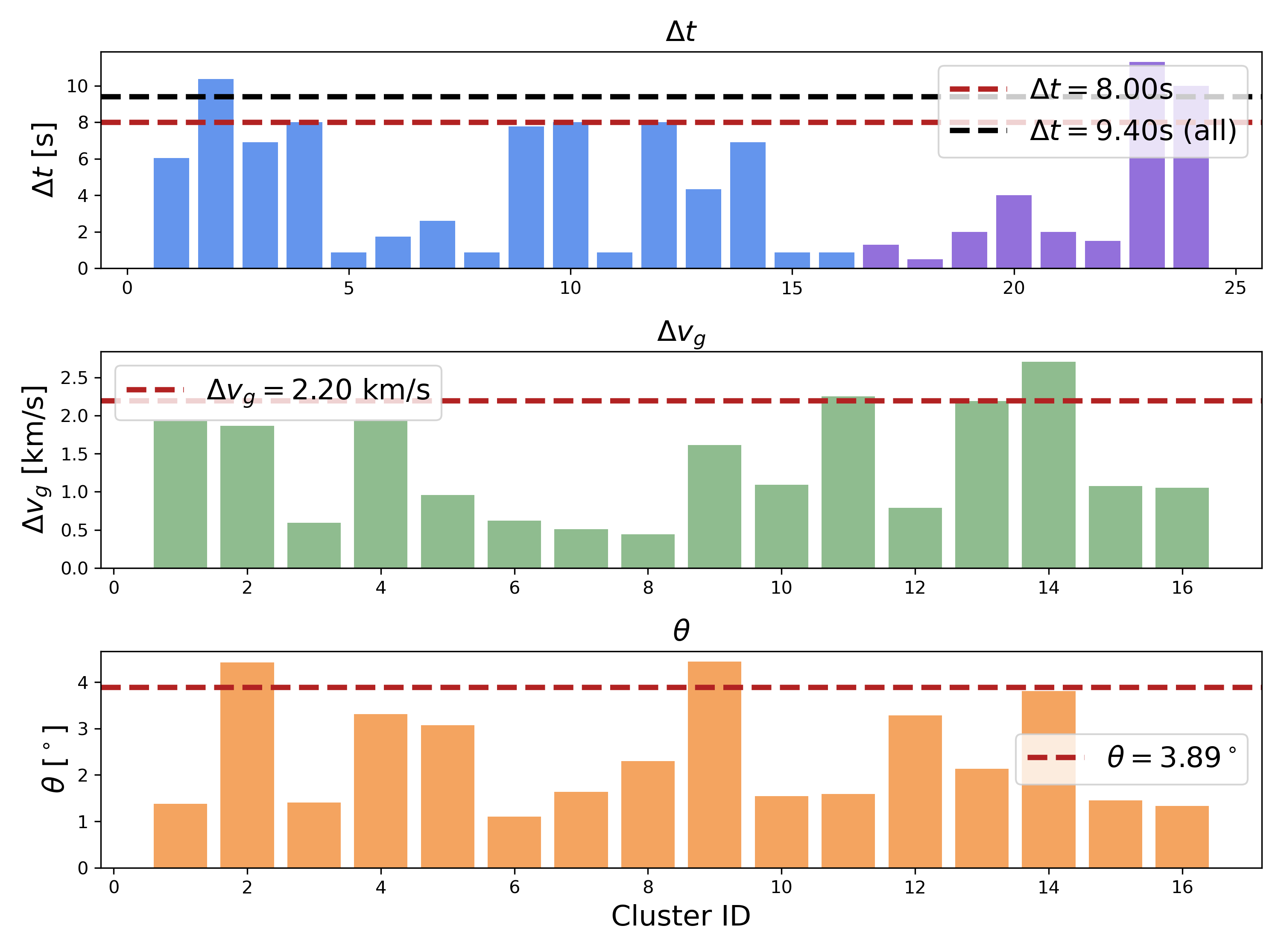}
    \caption{Distributions of the parameters of high-confidence clusters, where $\Delta t$ is the maximum time separation, $\Delta v_g$ is the maximum velocity separation between cluster fragments, and $\theta$ is the maximum angular separation between their geocentric radiants. The red dashed lines represent the $90^{th}$ percentiles of the dataset. In the topmost subplot, the blue bars represent the 16 clusters detected by us and the purple bars represent the 8 confirmed clusters; the black dashed line represents the $90^{th}$ percentile for the two sets combined.}
    \label{fig:parametersofrealclusters}
\end{figure}

To begin with, we examine the most critical parameter in the context of meteor clusters: the maximum time difference, $\Delta t$, between meteors within a cluster. Figure \ref{fig:parametersofrealclusters} (topmost subfigure) illustrates $\Delta t$ for the 16 high-confidence clusters we identified, as well as for the 8 confirmed clusters. Since there are no established thresholds for any parameter, we present the central tendency of the data by considering the $90^{th}$ percentile. For the 16 clusters, the $90^{th}$ percentile is 8.0 seconds and for the combined 24 clusters, it is 9.4 seconds. Here, we also know that the longest interval among the known meteor clusters was 11.3 seconds \citep{Vaubaillon2023_c7}. We find that all $\Delta t$ values lie within this range and are consistent with the conceptual definition of meteor clusters as short-duration events.

\begin{figure}[h]
    \centering
    \includegraphics[width=0.45\textwidth]{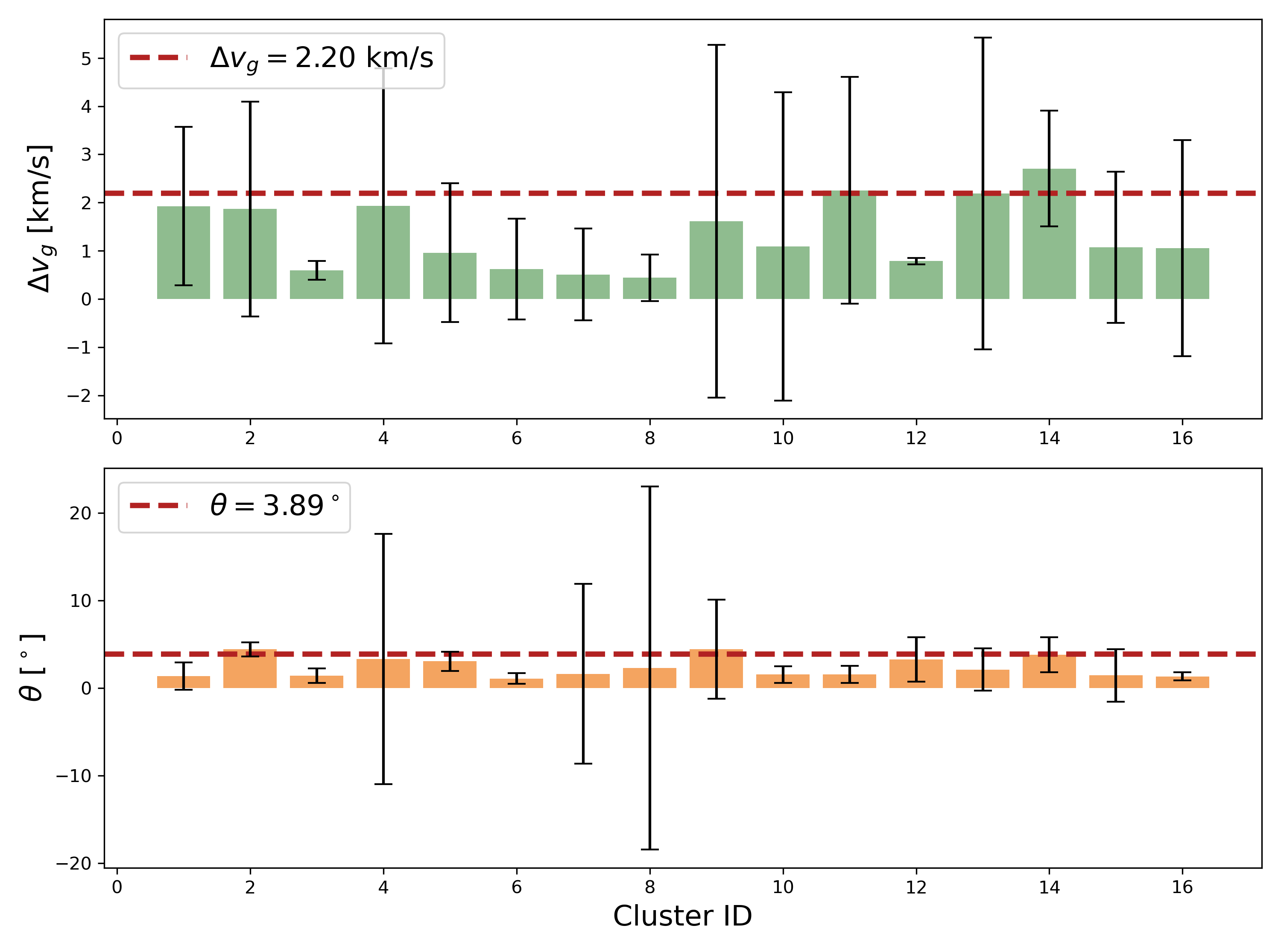}
    \caption{Uncertainties associated with $\Delta v_g$ and $\theta$.}
    \label{fig:uncertainties}
\end{figure}

Furthermore, the middle subfigure in Figure \ref{fig:parametersofrealclusters} shows the velocity difference, $\Delta v_g$, between the slowest and the fastest meteors within a cluster. In this case, a comparison between our high-confidence clusters and the eight confirmed clusters could not be performed as the information on velocities was not provided in all of the relevant literature. Consequently, among the 16 clusters considered, the $90^{th}$ percentile value for $\Delta v_g$ was found to be 2.20~km/s, with roughly half of the clusters exhibiting a $\Delta v_g$ of less than 1~km/s. It should be noted that a robust computation of meteor velocities from video and photographic records is an extremely complex process. \cite{Egal2017} showed that some data reduction pipelines introduce higher uncertainties on the velocity than claimed. 
The SonotaCo data reduction and orbit computation software is not publicly available, which prevents anyone from fully understanding the origin of the computed uncertainties.
The velocity discrepancies are evident in Figure \ref{fig:uncertainties} where the error bars are notably large in most cases. Such uncertainties are not optimal for our objective of defining the parameters of meteor clusters. This highlights the critical need for accurately determining orbital data using efficient pipelines.

The maximum angular separation, $\theta$, between the geocentric radiants of the cluster fragments is illustrated in the bottom subfigure of Figure \ref{fig:parametersofrealclusters}. Similar to $\Delta v_g$, data on the angular separation values for the eight detected clusters are absent in the literature, therefore, only the 16 that we identified are considered. For $\theta$, 90\% of the values lies below 3.89\textdegree. Similarly to $\Delta v_g$, we observe extremely large uncertainties for some data points, making it challenging to precisely define representative values of angular separation. 

Finally, the number of fragments within the high-confidence clusters ranges from four to seven, with the majority containing four meteors, which is relatively low compared to the confirmed clusters. Given that there are no defined limits for such characteristics of meteor clusters, it is challenging to ascertain what constitutes a low fragment count. Nevertheless, the clusters we have detected appear valid despite their lower fragment numbers. Thus, meteor clusters might not be as rare as only eight, but they might also not typically be as abundant as the majority of confirmed clusters. We believe that the previously detected clusters might have been detected primarily due to the higher intensity of their outbursts (i.e., higher number of fragments they contained) as well as their restricted physical extent, making them observable by a single camera.
These elements strongly suggest potential observational bias {for the detection of meteor clusters}.

While we present the central tendencies of the cluster candidates, defining strict threshold values may overly constrain what can be classified as a meteor cluster. Instead, we propose a working definition based on the probability of chance occurrence rather than fixed parameter values. Specifically and on the basis of fig. \ref{fig:probabilities}, if the expected number of random occurrences of such a cluster is less than 0.1, it can be considered a real cluster, as its occurrence is statistically unexpected. The relevant parameters, such as the number of meteors, time, velocity, and angular separation, can take any combination that results in this threshold being met.

\subsection{Future work and recommendations}

The developed method can be applied to any meteor orbit database, provided it includes the parameters described in sec. \ref{sec:param}.
Ideally, the camera's field of view, pointing direction and limited magnitude involved in the detection would also help to debias the observed number of clusters.
We would also recommend that future reports of meteor clusters include the number of fragments, duration of the event, difference of velocity and radiant separation, as it would help to define a cluster more precisely. In addition, we recommend that a detailed computation of the expected number of random occurrences of any future observed cluster is computed and published in order to confirm or revise our chosen threshold of 0.1. Finally, we recommend that the uncertainties of the angular separation and velocity are provided in order to strengthen the case of the detection of a meteor cluster.

Moreover, with the detection of more clusters than currently reported, and potential data from additional databases, the frequency of meteoroid self-fragmentation and its implications for the lifetime expectancy of meteoroids might be assessed.
Current models of lifetime expectancy represent the collisional lifetime of meteoroids, calculated from the characteristics of the meteoroid, as well as the number and speed of projectile particles in interplanetary space that are capable of collisionally destroying the target meteoroid \citep{Soja2019, Szalay2021}.
These collision models do not take into account the self-fragmentation lifetime of the meteoroids, which could potentially be a significant factor to consider.


\section{Conclusions}
After searching for meteor clusters in publicly available databases, we found 16 potential statistically significant clusters.
This is twice as much as currently reported (at the time this article is written).
Meteor clusters might be more frequent than currently reported. 
However, a fully debiasing analysis should take into account the atmospheric area surveyed by the cameras in order to definitively confirm the reality of the 16 potentially significant clusters.
Unfortunately, such debias is not possible given the current parameters available in databases.
We therefore recommend that anyone making a meteor orbit database includes the camera field of view, pointing direction and limiting magnitude available for each detection.
Any future meteor cluster detection claim should also provide a clear method to compute the expected number of cluster, to strengthen the case of a real cluster vs a random event. This number of expected events should be at most 0.1.
Nevertheless, this study helps characterize the most common features of meteor clusters, which may in turn help to constrain a future official definition of a meteor cluster.

\begin{acknowledgements}
The authors are thankful to P. Shober who greatly helped with advice regarding the choice and use of the cluster algorithm. We are also grateful to the people maintaining the IAU meteor orbit database, which proved useful for this exploratory work.
\end{acknowledgements}

%
%
\bibliographystyle{aa}
\bibliography{Biblio}  

\end{document}